\documentstyle[preprint,eqsecnum,prb,aps,tighten]{revtex}
\begin{document}
\preprint{ PUPT-94, UCSBTH-94}
%\draft

\def\ket#1{|#1\rangle}
\def\bra#1{\langle#1|}
\def\ve2{\vec{1\over 2}}
\newcommand{\be}{\begin{equation}}
\def\ee{\end{equation}}
\def\a{\alpha}
\def\b{\beta}
\def\ab{\bar{\alpha}}
\def\bb{\bar{\beta}}
\def\ib{\bar{i}}
\def\jb{\bar{j}}
\def\p1{ \psi_{L\alpha i}(z_1) }
\def\ra{\rightarrow}
\def\a{\alpha}
\def\b{\beta}l
\def\ab{\bar{\alpha}}
\def\bb{\bar{\beta}}
\def\ib{\bar{i}}
\def\jb{\bar{j}}
\def\be{\begin{equation}}
\def\ee{\end{equation}}

\centerline{cond-mat/yymmnn}
\vspace{1cm}

\centerline{\bf Majorana Fermions, Exact Mapping between Quantum Impurity}
\centerline{\bf  Fixed Points
 with four bulk Fermion species,}
\centerline{\bf and }
\centerline{ \bf Solution of the ``Unitarity Puzzle''}

\vskip 1cm
\centerline{\it Juan M. Maldacena${}^{1}$ and Andreas W.W.
Ludwig${}^{2,\dagger}$}

\vskip 1cm

\centerline{\it ${}^{1}$ Joseph Henry Laboratory of Physics,
Princeton University, Princeton, N.J. 08544}

\vskip .5cm
\centerline{\it ${}^{2}$  Physics Department, University of California,
Santa Barbara, Ca 93106}

\vskip 1cm
\centerline{\bf  Abstract }

Several Quantum Impurity problems with four flavors of bulk fermions
have zero temperature fixed points that show non fermi liquid
behavior. They include
the two channel Kondo effect, the
two impurity Kondo model, and the fixed point occurring
in the  four flavor Callan-Rubakov effect.
We provide a unified description which exploits the SO(8)
symmetry of the bulk fermions.
This leads
to a mapping between  correlation functions  of the different
models. Furthermore, we show that the two impurity Kondo fixed
point and the Callan-Rubakov fixed point are the same
theory.
All these models have the puzzling property that the
S matrix for scattering of fermions off the impurity
seems to be non unitary. We resolve this paradox
showing that the fermions scatter into collective
excitations which fit into the spinor representation of
SO(8). Enlarging the Hilbert space to include those
we find simple linear boundary conditions. Using these
boundary conditions
it is straightforward to recover all partition functions,
boundary states and correlation functions of these models.

\vskip 1cm
${}^{\dagger}$ A.P. Sloan Fellow

\vskip 10cm

\centerline{ \bf Introduction}

\vskip .5cm

Quantum Impurity problems  with four flavors of bulk fermions
such as    the 2-channel single impurity
and 1 channel two impurity  Kondo models have low temperature
fixed points that  provide
some of the simplest non-fermi liquid  theories.
Exact transport properties and correlation
functions have been computed for those fixed points
using conformal field theory (CFT) \cite{ludwig-fusion,%
ludwig-amplitude,ludwig-affleck-twoimp,affleck-saggi,ludwig-swaves,%
ludwig-affleck-green}.

The CFT treatment provides  exact
S-matrices for scattering of conduction electrons
off the impurity. In all cases mentioned above
we find that an incoming fermion that hits
the impurity seems to disappear because all
matrix elements with outgoing states in the
fermion Fock space are zero. We solve this
paradox by identifying the states into which
the fermions scatter as collective,
non local excitations.
In order to describe them it is useful to
view the four Dirac fermions as eight
Majorana fermions, and then notice the
presence of an SO(8) symmetry group.
The fermions transform in the vector representation
of SO(8). We will see that these excitations
will transform in a spinorial representation
which is also eight dimensional and has the
same charge, spin and flavor quantum numbers
of the electron.
Enlarging the Hilbert space to
include  these excitations,
we show that the boundary condition
is
linear
on the vector and the spinor
representations of $SO(8)$.  The vector is simply
scattered  into one of the spinors  with unit amplitude.
Using this linear boundary condition on spinors and vectors,
we recover all  CFT results including  partition functions, finite
size spectra, correlation functions,etc.. In this formalism
it is very clear how the original $SO(8)$ symmetry is
reduced to the symmetry groups of the various models.
Moreover, we show that the 2-impurity single channel non-fermi
liquid fixed point and the fixed point occurring
in the Callan-Rubakov effect are the same CFT, upon
an appropriate identification of the fermions.
Our interpretation in terms of $SO(8)$
thus provides a unified picture of all known
non-fermi liquid quantum impurity fixed points with
four bulk flavors and it leads to an explicit and exact
mapping between the theories.
As a consequence, we show that there are remarkable
and completely unexpected exact mappings
between the correlation functions
of those different models.

 Anisotropic versions
of the 2 channel Kondo and 2 impurity Kondo models
have also   been studied in the Toulouse limit \cite{emery-kivelson}
leading to results for some quantities in agreement
with the CFT treatment.
The Toulouse limit is fine tuned so that the
leading irrelevant operator is absent, giving rise to some
ungeneric features. Also, correlation functions and
transport properties have not been computed using
the Toulouse  limit. Our $SO(8)$ Majorana fermion
 treatment is very much like an
exact and {\it isotropic}  version of the Toulouse limit approximation.
It incorporates the leading irrelevant operator, and
permits to compute exactly all correlation functions,
in agreement with previously obtained CFT results.

In section I we review the various models we will analyze. In section
II we describe the fermions in terms of SO(8) representations, we
introduce the spinors and we define a new set of fermions which
have simple boundary condition at all fixed points.
We then analyze the various impurity models using this
formalism, showing the connection between them as well
as explaining the resolution to the unitarity problem.
In section III we calculate correlation functions
for the Kondo fixed using the new formalism. We study
four point correlators, and the leading irrelevant
correction to the
two point functions.
In section IV we show how the symmetry groups of the
various models arise from the original SO(8) symmetry,
and we provide the mapping between correlation functions.
In appendix A we analize more explictly the
relationship between Kondo and Monopole correlators.
And in appendix B we compute explicitly the boundary
states and partition functions.

\section{Review of the various impurity models}

We will be considering in this paper the critical points of
various impurity models. We start with a conformal invariant
 bulk Hamiltonian
describing  four Dirac  (eight Majorana) free fermions.
Then,  a boundary
 interaction with a localized degree of freedom is introduced.
Conformal invariance is
broken, but, in the infrared limit, the theory  flows to a boundary fixed
point that is conformal invariant. We will study these nontrivial fixed
points.

The first model that we will consider is the spin 1/2  two channel Kondo
effect, which consists of a spin 1/2 impurity interacting with the spin
of two channels (flavors) of electrons.
Putting the impurity  $\vec S$ at
the origin of three dimensional space, we see that
it interacts only with the
s-wave. The problem can be reduced  to a 1+1 dimensional theory with
a boundary at $x=0$ where the spin sits.
 The four Dirac fermions correspond to the spin up and
spin down components of the two channels  of three dimensional
electrons.
The Hamiltonian is
$$
H= { v_F \over 2 \pi } \int_0^\infty dr (\psi^{\dagger \alpha j}_L(r) i
{d \psi_{L \alpha j }(r) \over dr} -
\psi^{\dagger \alpha j}_R(r) i
{d \psi_{R \alpha j }(r) \over dr} )  + v_F \lambda_K {\vec J}_L(r=0) . {\vec
S}
$$
The first term describes the four Dirac fermions $\psi_{\alpha j}$, where
$\alpha $ is the spin index and $j$ is the flavor index. The second term
describes the interaction of the impurity spin with the total fermion
spin current
$j^a = \psi^{\dagger \alpha j} (\sigma^a)^{\,\beta}_\alpha \psi_{\beta j}$
at the origin.
 It turns out that  under the renormalization group
this theory flows in the infrared limit to
a new critical point at a finite value of $\lambda_K$. At this critical
point the theory is conformal invariant, leading to an infinite number
of conservation laws. The charge, spin and flavor symmetries
appear  in conformal field theory as a
U(1)$^{charge}\times$SU(2)$^{spin}_2\times$SU(2)$^{flavor}_2$
 Kac-Moody algebra.  The constraints imposed by this algebra
lead to a complete classification of the possible theories, one for
each value of the spin of the impurity $ 0 \leq s \leq  1 $.
In the case  $s=1/2$ we get a nontrivial fixed point that
 was extensively analyzed  using
the fusion hypothesis \cite{ludwig-fusion}. This procedure consists in imposing
an auxiliary boundary condition
at $x=l$ and then studying  the partition function for the system at
finite temperature $T = 1/\beta$. The partition function has  the form
$$ Z = Tr( e^{-\beta H } ) = \sum_i n_i \chi_i ( e^{-\beta/l} ) $$
Where $i \equiv ( Q,j_{sp},j_{fl}) $
 runs over the primary fields of the theory which are labeled
by charge, spin and flavor quantum numbers with $\chi_i$ the corresponding
character.  $n_i$ indicates the
number of times that this primary field appears in the partition function.
These numbers, that  depends on the boundary conditions at both ends, have
to obey certain constraints due to conformal invariance \cite{cardy} .
The fusion procedure provides a way to generate solutions to these constraints.
Starting with the primary field content of the theory with free fermion
boundary conditions ($ n_i^{FF}$), we calculate the primary field content with
the formula
$
 n_i^{KF} =
 N^k_{ij} n_k^{FF} $, where
$N^k_{ij}$ are the fusion coefficients for the fields $i$ and $j$.
The partition function with a Kondo boundary is obtained fusing with
a spin 1/2 operator (i.e. $j=(0,1/2,0)$).

An important tool in the analysis of these problems was introduced by
Cardy \cite{cardy}. It is based on the observation that we can
view this system as  propagating in imaginary time with
$Z = Tr( e^{-\beta H_{kondo}}) $ or as propagating
in space between two boundary states so that
$Z= \bra K e^{-l H_{free} } \ket F$. In this later
picture all the information about the boundary interaction is encoded
in  the Kondo boundary state $\ket K$ and we have just a free Hamiltonian.

It is interesting to analyze the scattering of fermions at the Kondo
boundary. We send in a left moving fermion and we try to find out
what the outgoing excitations are.
This scattering amplitude is related  to  the
correlation function of a left moving with a right moving fermion which is
constrained  by conformal invariance to be of the form
$$ \langle \psi^{\dagger \alpha j}(z) \psi_{ \alpha j }({\bar z}) \rangle=
{S_1 \over (z -{\bar z})^2 }
$$
The scattering amplitude  $S_1$ was shown  in \cite{ludwig-amplitude}
to be $ S_1 =0 $. Furthermore, the scattering amplitudes
for a fermion to go into any number of fermions is zero so that
unitarity seems to be violated.
We will address this paradox below.

The second model we will analyze is the four flavor Callan-Rubakov
effect which describes
the scattering  of
 fermions by a magnetic monopole \cite{polchinsky,callan-rubakov}.
This scattering has the surprising feature of catalyzing baryon
number violation.
In the low energy limit the theory will flow to
a conformal field theory having
an
SU(4) Kac-Moody
algebra  associated with flavor conservation. The free fermion theory
contains also a  U(1) current corresponding to baryon number, which
is not conserved  by  the interaction with the monopole.
It was found in
\cite{polchinsky,affleck-saggi} that  the low energy theory
reduces  to a simple  change in the  boundary condition for this current to
\begin{equation}
j_L^{U(1)}(z) = - j_R^{U(1)}({\bar z})|_{Im z =0}
\label{boundaryuonef}
\end{equation}
This  conformal point was described in \cite{affleck-saggi} using
a U(1)$\times$SU(4) decomposition of the free fermion theory and
then twisting the U(1) boson as implied in \ref{boundaryuonef}.
The gauge theory that produces the monopole
has a topological
$\theta$ angle which  appears  as parameter of the low energy theory.
It introduces
an additional phase in the scattering amplitudes.
When we study the scattering of a single fermion by the monopole
we
also see that
unitarity seems to be violated because the
correlation functions of a left moving fermion with any number of
right moving fermions is zero.

The last model we will discuss is the one channel two impurity spin
1/2 Kondo effect. This model is a step forward in analyzing the effect
that interimpurity couplings produce in a realistic system. It consists
of two spin 1/2 impurities located a two different points interacting
with the spin of one channel (one band) of electrons at those points.
It is possible to reduce it to a 1+1 dimensional problem. In the
reduction process one needs to introduce two channels
of 1+1 dimensional fermions, which together with two spin
directions give  four bulk fermions.
 The boundary interaction contains several parameters
associated with the interaction of these two channels with both impurities,
together with
an interimpurity coupling of the form $K {\vec S}_1 \cdot {\vec S}_2$.
For $K \rightarrow \infty$ the impurity forms a singlet and the
theory reduces in the infrared to the free fermion theory. If
$K \rightarrow - \infty$ the impurity forms a spin one triplet and the
fixed point is the same as the one for the  spin one two channel Kondo theory,
which is a fermi liquid  with a phase shift of $\pi$ at the origin.
As the phase shift for the fermions
 can only be zero or $\pi$ for a particle-hole symmetric
Hamiltonian there is a non fermi liquid fixed point for a critical value
of the coupling $K_c$.
This intermediate fixed point was analyzed by \cite{ludwig-affleck-twoimp}.
For the critical value of the parameters the charges of the two
channels are conserved independently. Associated with each of
these U(1) currents there is a boson with a compactification radius such
that the U(1) symmetry is enhanced to SU(2)$_1$.
The resulting conformal structure is
 SU(2)$_1\times$SU(2)$_1\times$SU(2)$_2\times$Ising.
The first
two SU(2)s correspond to the charges of both channels. The third SU(2) is
associated with the spin and it is also
 necessary to introduce an extra Ising model
degree of freedom. It was found in \cite{ludwig-affleck-twoimp} that
the theory is described just by a change in the boundary condition
for the Ising model fermion. The same unitarity paradox
we had for the previous cases is present here,
 namely that a left moving fermion has no
overlap with right moving fermions. It was also found that the
theory has actually a hidden SO(7) invariance.

\section{ New  description using SO(8) }
\subsection{ Free Fermions: $SO(8)$ and its representations }

 We will concentrate now on the free fermion theory and consider only
the left moving sector. All we will say goes  over to the right moving
sector in a straightforward way.
 We have four species of Dirac fermions $ \psi_{\alpha j}$ labeled by the
spin $\alpha =1,2$ (for $ \uparrow, \downarrow $) and the flavor index $j
=1,2$.
We can form eight Majorana fermions $\chi_{a}, a =1,2...,8$ by taking the real
and imaginary parts of the complex dirac fermions. The Hamiltonian is
\begin{equation}
H^0_L = {1 \over 2\pi} \int dx \chi_{a} i { d \over dx } \chi_{a}
\end{equation}
which is invariant under SO(8) rotations of the fermions generated by the
currents
\begin{equation}
j^A(z)= \chi_{a}(z) (T^A)_{a b} \chi_{b}(z)
\label{so8currents}\end{equation}
where $T^A$ are antisymmetric 8$\times$8  associated with the
generators of SO(8).
These currents form a Kac-Moody algebra at level one which has four
representations \cite{georgi-slansky}:
 a 1  dimensional singlet $(1)$,
an 8 dimensional vector $(v)$, and two irreducible spinor representations,
$(s)$ and $(c)$, each of dimension 8.
The one dimensional singlet corresponds to the identity operator and the
vector $(v)$ to the fermions.
There are  two sectors in the free fermion Hilbert space which
arise when we consider
different boundary conditions in the space direction
\begin{eqnarray}
{\rm NS}_{x}:  \chi_a(\tau,x+l)=- \chi_a(\tau,x) ~~~~\mbox{Neveu-Schwartz}\\
{\rm R}_{x}:  \chi_a(\tau,x+l)= \chi_a(\tau,x)~~~~\mbox{Ramond}
\label{NSRx}
\end{eqnarray}
The spinors appear in the Ramond sector, in fact
we can decompose  the two sectors in terms of representations as
 \begin{equation}
{\rm NS} = (1) + (v), \qquad {\rm R} = (s) + (c)
\end{equation}
We note, for later usage,
 that if we study the system at finite temperature, considering the
system in periodic euclidean time $ t= t +\beta$, we have also two
possible boundary conditions
\begin{eqnarray}
{\rm NS}_{\tau}:  \chi_a(\tau+\beta,x)=- \chi_a(\tau,x)\\
{\rm R}_{\tau}:  \chi_a(\tau+\beta,x)= \chi_a(\tau,x)
\label{NSRt}
\end{eqnarray}
In usual physical situations one considers only the antiperiodic (NS) sector
in time, this is the one that appears naturally when we perform the trace
over the fermions to get the partition function.
Note that for each sector the boundary condition is the same for all eight
fermions, as required by SO(8) invariance.
In order to describe the spinors it is useful to bosonize the theory
introducing
 four left moving bosonic fields
 \begin{eqnarray}
:\psi^{\dagger\alpha j}\psi_{\alpha j}: (z)= i\partial_z \phi_{\alpha j}
\qquad ({\rm no \ summation})
\label{currentsold} \end{eqnarray}
These currents  are of the form \ref{so8currents} and
correspond to  four commuting elements of the SO(8)
algebra, which is the  maximum number of
commuting generators we can have because the algebra
 has rank four. They are conventionally called
{\it Cartan generators} and will be denoted by $H^1,...,H^4$.
In terms of these bosons we can write the fermions as \footnote{
Actually, we have to insert additional operators in front of the
exponentials to ensure proper anticommutation relations. These are
just constants and we will omit them for the sake of clarity.}
\begin{eqnarray}
\psi_{\alpha j} =  e^{-i \phi_{\alpha j}},\qquad
\psi^{\dagger \alpha j} =  e^{i \phi_{\alpha j}}
\label{abelian1}
\end{eqnarray}
The bosons satisfy the periodicity condition
\begin{equation}
\phi_{\alpha j}(z)=\phi_{\alpha j}(z)+2 \pi
\label{periodicity1}
\end{equation}
which in SO(8) language it just means that a rotation
by $ 2 \pi$ leaves the system invariant. The spinors
however are expected to aquire a minus sign under a
$ 2 \pi$ rotation. We also know that the spinors
have conformal dimension 1/2. These requirements fix
the spinors to be
\begin{eqnarray}
c_{\mu}(z) \equiv \left \{
\begin{array}{l}
e^{{i \over 2}( \phi_{1,1}+\phi_{2,1}+\phi_{1,2}+\phi_{2,2})} \\
e^{{i \over 2}( \phi_{1,1}-\phi_{2,1}+\phi_{1,2}-\phi_{2,2})} \\
e^{{i \over 2}( \phi_{1,1}+\phi_{2,1}-\phi_{1,2}-\phi_{2,2})} \\
e^{{i \over 2}( \phi_{1,1}-\phi_{2,1}-\phi_{1,2}+\phi_{2,2})} \\
e^{-{i \over 2}( \phi_{1,1}+\phi_{2,1}+\phi_{1,2}+\phi_{2,2})} \\
e^{-{i \over 2}( \phi_{1,1}-\phi_{2,1}+\phi_{1,2}-\phi_{2,2})} \\
e^{-{i \over 2}( \phi_{1,1}+\phi_{2,1}-\phi_{1,2}-\phi_{2,2})} \\
e^{-{i \over 2}( \phi_{1,1}-\phi_{2,1}-\phi_{1,2}+\phi_{2,2})} \\
\end{array}\right \}, \quad
s_{\mu}(z) \equiv \left \{
\begin{array}{l}
e^{{i \over 2}( \phi_{1,1}+\phi_{2,1}+\phi_{1,2}-\phi_{2,2})} \\
e^{{i \over 2}( \phi_{1,1}+\phi_{2,1}-\phi_{1,2}+\phi_{2,2})} \\
e^{{i \over 2}( \phi_{1,1}-\phi_{2,1}+\phi_{1,2}+\phi_{2,2})} \\
e^{{i \over 2}( -\phi_{1,1}+\phi_{2,1}+\phi_{1,2}+\phi_{2,2})} \\
e^{-{i \over 2}( \phi_{1,1}+\phi_{2,1}+\phi_{1,2}+\phi_{2,2})} \\
e^{-{i \over 2}( \phi_{1,1}+\phi_{2,1}-\phi_{1,2}+\phi_{2,2})} \\
e^{-{i \over 2}( \phi_{1,1}-\phi_{2,1}+\phi_{1,2}+\phi_{2,2})} \\
e^{-{i \over 2}( -\phi_{1,1}+\phi_{2,1}+\phi_{1,2}+\phi_{2,2})}
\end{array}\right \}
\label{spinor1}
\end{eqnarray}
We have disentangled the two irreducible spinor representations
using the chirality operator $\Gamma$ which commutes with
SO(8) and satisfies $\Gamma ^2 =1 $. It is represented in our case
by $(-1)^{2 Q}$ where $Q$ is the total charge corresponding to
the operator
\begin{eqnarray}
j^{ch}(z) \equiv {1 \over 2}
\sum_{\alpha,j=1,2} :\psi^{\dagger \alpha j}\psi_{\alpha j}: (z) =
i \partial_z \phi^{ch}(z),\\
{\rm where} \quad  \phi^{ch}(z) \equiv {1 \over 2}
\sum_{\alpha,j=1,2} \phi_{\alpha j}(z)
\label{chargecurrent}
\end{eqnarray}
In order to simplify later formulas we
have adopted a slightly unconventional
normalization for the charge operator that assigns
charges $\pm 1/2 $ for the fermions \ref{abelian1}.
We  introduce  also bosonic fields associated to the
third component of the spin and the third component of flavor
\begin{eqnarray}
j^{sp} (z)& =&
 {1 \over  2} :\psi^{\dagger \alpha j} (\sigma^z)^{\beta}_{\alpha}
\psi_{\beta j}: (z)
 =i \partial_z \phi^{sp}(z)={ 1\over 2 }
\sum_{\alpha,j=1,2} (\sigma^z)^{\alpha}_{\alpha}
i\partial_z \phi_{\alpha,j}(z) \nonumber \\
j^{fl}(z)& = &{1 \over  2} :\psi^{\dagger \alpha i} (\tau^z)^{j}_{i}
\psi_{\alpha j}: (z)
 =i \partial_z \phi^{fl}(z)={ 1\over 2 }
\sum_{\alpha,j=1,2} (\tau^z)^j_j
i\partial_z \phi_{\alpha,j}(z)
\label{newcurrents}
\end{eqnarray}
which together with \ref{chargecurrent} form a set of three commuting
generators. By orthogonality we  find a fourth one
\begin{eqnarray}
j^{X}(z)&=&
{1 \over  2} :\psi^{\dagger \alpha i}
( \sigma^z)^{\beta}_{\alpha}(\tau^z)^{j}_{i}
\psi_{\beta j}: (z)
 =i \partial_z \phi^{X}(z)={ 1\over 2}
\sum_{\alpha,j=1,2} (\sigma^z)^{\alpha}_{\alpha}
(\tau^z)^j_j i\partial_z \phi_{\alpha j}(z)
\label{xboson}
\end{eqnarray}
These four bosons constitute a
new set of Cartan generators. They are related to
to the old ones  by
\begin{eqnarray}
J^{ch}_0 = {\tilde H}^1 = {1 \over 2}  ( H^1 + H^2 + H^3 + H^4) \nonumber \\
J^{sp}_0 = {\tilde H}^2 = {1 \over 2}  ( H^1 - H^2 + H^3 - H^4) \nonumber \\
J^{fl}_0 = {\tilde H}^3 = {1 \over 2}  ( H^1 + H^2 - H^3 - H^4) \nonumber \\
J^{X}_0 = {\tilde H}^4 = {1 \over 2}  ( H^1 - H^2 - H^3 + H^4)
\label{newcartan}
\end{eqnarray}
which is the relation between the old bosons \ref{abelian1}
and the new bosons $\phi^{ch},~ \phi^{sp},~ \phi^{fl},~ \phi^{X}$.
We  rewrite now  the vector and spinor representations as
\begin{equation}
c_{\mu}(z) \equiv \left \{
\begin{array}{l}
e^{i \phi^{ch}}\\
e^{i \phi^{sp}}\\
e^{i \phi^{fl}}\\
e^{i \phi^{X}}\\
e^{-i \phi^{ch}}\\
e^{-i \phi^{sp}}\\
e^{-i \phi^{fl}}\\
e^{-i \phi^{X}}
\end{array}\right \}, \
\left \{
\begin{array}{l}
\psi^{\dagger,1,1} \\
\psi^{\dagger,2,1} \\
\psi^{\dagger,1,2} \\
\psi^{\dagger,2,2} \\
\psi_{1,1} \\
\psi_{2,1} \\
\psi_{1,2} \\
\psi_{2,2}
\end{array}
\right \}=
 \left \{
\begin{array}{l}
e^{{i \over 2}( \phi^{ch}+ \phi^{sp}+ \phi^{fl}+ \phi^{X}) }\\
e^{{i \over 2}( \phi^{ch}- \phi^{sp}+ \phi^{fl}- \phi^{X}) }\\
e^{{i \over 2}( \phi^{ch}+ \phi^{sp}- \phi^{fl}- \phi^{X}) }\\
e^{{i \over 2}( \phi^{ch}- \phi^{sp}- \phi^{fl}+ \phi^{X}) }\\
e^{-{i \over 2}( \phi^{ch}+ \phi^{sp}+ \phi^{fl}+ \phi^{X}) }\\
e^{-{i \over 2}( \phi^{ch}- \phi^{sp}+ \phi^{fl}- \phi^{X}) }\\
e^{-{i \over 2}( \phi^{ch}+ \phi^{sp}- \phi^{fl}- \phi^{X}) }\\
e^{-{i \over 2}( \phi^{ch}- \phi^{sp}- \phi^{fl}+ \phi^{X}) }\\
\end{array}\right \},
\
s_{\mu}(z) =
 \left \{
\begin{array}{l}
e^{{i \over 2}( \phi^{ch}+ \phi^{sp}+ \phi^{fl}- \phi^{X}) }\\
e^{{i \over 2}( \phi^{ch}- \phi^{sp}+ \phi^{fl}+ \phi^{X}) }\\
e^{{i \over 2}( \phi^{ch}+ \phi^{sp}- \phi^{fl}+ \phi^{X}) }\\
e^{{i \over 2}( \phi^{ch}-\phi^{sp}- \phi^{fl}- \phi^{X}) }\\
e^{-{i \over 2}( \phi^{ch}+ \phi^{sp}+ \phi^{fl}- \phi^{X}) }\\
e^{-{i \over 2}( \phi^{ch}- \phi^{sp}+ \phi^{fl}+ \phi^{X}) }\\
e^{-{i \over 2}( \phi^{ch}+ \phi^{sp}- \phi^{fl}+ \phi^{X}) }\\
e^{-{i \over 2}( \phi^{ch}-\phi^{sp}- \phi^{fl}- \phi^{X}) }\\
\end{array}\right \}
\label{newspinorsvector}
\end{equation}

We see that, in terms of the new bosons, the vector $(v)$ has taken the
form of the spinor and the spinor $(c)$ has taken the form of the vector.
This surprising property of SO(8), namely
that we can interchange the 8 dimensional representations by
choosing a different basis of Cartan generators,
is called triality and leads to a very simple and unified
picture for the impurity problems.
Clearly we cannot have changed the physics  when we  chose this
new basis of Cartan generators so  the correlation functions
are the same, using either basis.
An important point in understanding this triality operation is that
in the original basis the eigenvalues of the Cartan generators
$H^1,\cdots, H^4$ for the vector representation
are integers (such that the sum is an odd number).
By means of \ref{newcartan} we see that the new generators have
half integer eigenvalues (such that  the sum is  an even number).

Finally we like to note that the last two rep's in \ref{newspinorsvector}
differ {\it only} in the sign of the boson $\phi^X$. This
will become a crucial observation later on.

We  introduce now  a new set of Dirac fermions
\begin{equation}
\psi_{ch} = e^{- i \phi^{ch} }, ~~~~~~~~
\psi_{sp} = e^{- i \phi^{sp} }, ~~~~~~~~
\psi_{fl} = e^{- i \phi^{fl} }, ~~~~~~~~
\psi_{X} = e^{- i \phi^{X} }
\label{newfermions}
\end{equation}
which form  the components of the $(c)$ spinor \ref{newspinorsvector}.
The currents \ref{chargecurrent} \ref{newcurrents} have the following
form
\begin{equation}
j^{ch} = :\psi^{\dagger ch}  \psi_{ch}: ~~~~~~~~~~~~
j^{sp \,3} = :\psi^{\dagger,sp} \psi_{sp}:~~~~~~~~~~~~
j^{fl \, 3} = :\psi^{\dagger,fl} \psi_{fl}:
\end{equation}
in terms of these fermions.
We  introduce eight Majorana fermions
 by taking the
real and imaginary parts of the Dirac fermions \ref{newfermions}
\begin{eqnarray}
\chi^{A}_1 = { \psi^{\dagger A} + \psi_{A} \over 2} =\cos \phi^A
\nonumber \\
\chi^{A}_2 = { \psi^{\dagger A} - \psi_{A} \over 2 i}= \sin \phi^A
\label{majorana}
\end{eqnarray}
where the label $A$ runs over the four fermions \ref{newfermions}

In order to clarify further the connection with the original
SU(2)$\times$SU(2)$\times$U(1) description  we
build the SU(2) groups from these fermions. The SU(2)$_2$ algebra
can be realized with three Majorana fermions.
So $ \chi^{sp}_1,~ \chi^{sp}_2 $ together with
 $\chi^{X}_2$ gives rise to the SU(2)$^{spin}$ group.
In a similar fashion $\chi^{X}_1$ combines with the other two
flavor fermions to give SU(2)$^{flavor}$. So that
the other components of the SU(2) currents read
\begin{eqnarray}
%j^{sp \, +} = \psi^{\dagger,sp}{1\over 2 i}  ( \psi^{\dagger,x}- \psi_{X}),
%\qquad
%j^{sp \, -} = \psi_{sp} {1\over 2 i}  ( \psi^{\dagger,x}- \psi_{X} ) \\
%j^{fl \, +} = \psi^{\dagger,fl}{1\over 2}  ( \psi^{\dagger,x}+ \psi_{X}),
%\qquad
%j^{fl \, -} = \psi_{fl} {1\over 2}  ( \psi^{\dagger,x}+ \psi_{X} )
%
j^{sp \, +} = \psi^{\dagger,sp} \chi^X_2= e^{i \phi^{sp}} \sin \phi^X ,
\qquad
j^{sp \, -} = \psi_{sp} \chi^X_2= e^{ -i \phi^{sp}} \sin \phi^X
\label{currentspmspin}
 \\
j^{fl \, +} = \psi^{\dagger,fl}\chi^X_1= e^{i \phi^{fl}} \cos \phi^X,
\qquad
j^{fl \, -} = \psi_{fl} \chi^X_1= e^{ -i \phi^{fl}} \cos \phi^X
\label{currentspm}
\end{eqnarray}

At these point we have two sets of fermions (and the corresponding bosons)
 related to each other in a nonlocal way that correspond to two ways
of choosing Cartan generators for the SO(8) symmetry algebra.
The importance of this construction is  that while the boundary
condition is nonlinear in terms of the original fermions $\psi_{\alpha j}$
it  is indeed linear in the new fermions \ref{newfermions} as we will see
below.

\subsection{Spin 1/2 two channel Kondo fixed point}

The fixed point is described by a boundary conformal field theory.
The  SU(2)$\times$SU(2)$\times$U(1) symmetry of the Kondo model, together
with conformal invariance is
 preserved at the boundary.  This means that the currents must satisfy
the boundary condition
\begin{equation}
j^a_L(z) = j^a_R({\bar z})|_{Im z =0}
\label{currentcons}
\end{equation}
 The operators in \ref{chargecurrent}
\ref{newcurrents} are just some of these currents, this  implies that
the left and right components of the bosons $\phi^{ch}, \phi^{fl}\phi^{sp}
$ become equal at the boundary, up to a possible constant that
will be determined later.
We will determine the boundary condition of the boson $\phi^{X}$ in
two ways. First let us
consider
scattering off the boundary. It was found in \cite{ludwig-amplitude}
 that a left moving fermion
$\psi_{L \alpha j}(z)$
has zero probability to scatter into a right moving fermion.
The fermion is a primary field under the
Kac Moody symmetry and there are only a finite number of primary fields
into which the fermions can scatter. The only one that has the same
SU(2)$\times$SU(2)$\times$U(1) quantum numbers  is the
$(s)$ spinor. A glance at \ref{newspinorsvector} convinces us that
the boundary condition is
\begin{equation}
\phi_L^X(z) = - \phi^X_R({\bar z})|_{Im z=0}
\label{boundaryx}
\end{equation}
We see therefore that this twist of the $x$-boson is equivalent to
the the fact that the left moving fermions have zero overlap with
right moving fermions.
The spinors appear in the fermionic theory as collective
excitations. The appearance of solitons upon scattering
from a boundary was also seen in the simpler case
of a scalar field in \cite{cklm}.

The second way of deducing this boundary condition is based in
the observation that $j^X=i\partial \phi^X $ is the
(3,3) component of a flavor-spin one primary field
($(Q,j^{sp},j^{fl})=(0,1,1)$).
The correlation functions of left and right moving fields
of this type, calculated in
\cite{ludwig-amplitude}, show
 the presence of a
minus sign upon reflection at the boundary. This leads to the
boundary condition \ref{boundaryx}.

Demanding that the other two components of the flavor and spin currents
\ref{currentspm} are conserved we find
\begin{equation}
\phi_L^{fl}(z) = \phi^{fl}_R({\bar z})
 |_{Im z=0}~~~~~~~~~~~~~
\phi_L^{sp}(z) = \phi^{sp}_R({\bar z}) + \pi|_{Im z=0}
\label{boundaryspfl}
\end{equation}
so we fixed some of the constants we mentioned above. The charge
boson will, in general, involve an arbitrary constant that will appear
as an extra phase in the scattering amplitudes. We will set it to
zero putting
\begin{equation}
\phi_L^{ch}(z) = \phi^{ch}_R({\bar z})|_{Im z=0}
\label{boundarych}
\end{equation}
This corresponds to the particle hole symmetric case. We will show that
this model has enhanced symmetry.

In terms of the fermions \ref{newfermions} the boundary
conditions are linear
\begin{equation}
\psi_{L,ch}= \psi_{R,ch},\qquad \psi_{L,fl}= \psi_{R,fl}
\qquad \psi_{L,sp}= - \psi_{R,sp},\qquad
\psi_{L,x}=  \psi_{R}^{\dagger,x}
\label{kondoboundary}
\end{equation}
while in terms of the original fermions they read
\begin{equation}
 \psi_{L\alpha i}(\bar{z}) = e^{i\pi j_\alpha } S_{R\alpha
i}(\bar{z})~~~~~~~~~~~~~~
S_{L \alpha i}(\bar{z})
= e^{i \pi j_\alpha } \psi_{ R \alpha i}(\bar{z})
\label{kondoboundfer}
\end{equation}
where $j_\alpha = \pm {1 \over 2 } $ is the spin quantum number of the
operator.

This is a remarkably simple picture of the boundary condition:
 the three Majorana fermions associated with the
SU(2)$^{spin}$ group aquire a minus sign. In this picture it
is clear that the original SO(8) symmetry is broken down to
SO(3)$\times$SO(5). The group SO(5) contains the  expected
SU(2)$^{flavor}\times$U$^{charge}$(1) plus the following extra generators
\begin{equation}
 \cos \phi^X e^{\pm i \phi^{ch} } ~~~~~~~~~ e^{\pm i \phi^{ch} \pm i \phi^{fl}}
\end{equation}
which transform particles to holes. Therefore  this symmetry will only
be present in the particle hole symmetric case. Indeed, it is easy to see that
it is broken if  we include a constant in the boundary condition
for $\phi^{ch}$ ($ \phi^{ch}_L= \phi^{ch}_R + \delta|_{Im z =0} $ ).

Using the bosonic representation \ref{newspinorsvector} and the
boundary conditions \ref{boundaryx} \ref{boundaryspfl} \ref{boundarych}
the problem of calculating correlation functions in this theory is
reduced to a simple
free field theory exercise. Using this bosons we can also
write explicitly the Kondo boundary state and calculate
the partion function of the theory. We leave these calculations for
Appendix B, but let us anticipate that they will agree with the
results found using the fusion method \cite{ludwig-fusion}

\subsection{Monopole fixed point}

The Monopole theory is described in terms of four fermion flavors
$ \Psi_1,...,\Psi_4 $, the natural
 U(1)$\times$SU(4) group is broken to
SU(4) due to a change in the boundary condition for the U(1)
baryon number current
to
\begin{equation}
j^{U(1)}_L(z) = - j^{U(1)}_R({\bar z})|_{Im z =0}
\label{boundaryuone}
\end{equation}
This reminds us of the condition
\ref{boundaryx} for  $\phi^X$.
So it is natural to make a correspondence between the fermions
$\Psi_1,...,\Psi_4$ of the monopole theory with the ones of
the Kondo theory in such a way that the baryon number
 becomes the U(1)$^X$ charge of the Kondo theory
\begin{equation}
\left \{
\begin{array}{l}
\Psi^{\dagger,1} \\
 \Psi_{2} \\
 \Psi_{3} \\
\Psi^{\dagger,4} \\
\Psi_{1} \\
\Psi^{\dagger,2} \\
 \Psi^{\dagger,3} \\
 \Psi_{4} \\
\end{array}
\right \}=
\left \{
\begin{array}{l}
\psi^{\dagger,1,1} \\
\psi^{\dagger,2,1} \\
\psi^{\dagger,1,2} \\
\psi^{\dagger,2,2} \\
\psi_{1,1} \\
\psi_{2,1} \\
\psi_{1,2} \\
 \psi_{2,2}
\end{array}
\right \}=
 \left \{
\begin{array}{l}
 e^{{i \over 2}( \phi^{ch}+ \phi^{sp}+ \phi^{fl}+ \phi^{X}) }\\
e^{{i \over 2}( \phi^{ch}- \phi^{sp}+ \phi^{fl}- \phi^{X}) }\\
e^{{i \over 2}( \phi^{ch}+ \phi^{sp}-\phi^{fl}- \phi^{X}) }\\
e^{{i \over 2}( \phi^{ch}- \phi^{sp}-\phi^{fl}+ \phi^{X}) }\\
e^{-{i \over 2}( \phi^{ch}+ \phi^{sp}+\phi^{fl}+ \phi^{X}) }\\
e^{-{i \over 2}( \phi^{ch}- \phi^{sp}+\phi^{fl}- \phi^{X}) }\\
e^{-{i \over 2}( \phi^{ch}+ \phi^{sp}- \phi^{fl}- \phi^{X}) }\\
e^{-{i \over 2}( \phi^{ch}- \phi^{sp}- \phi^{fl}+ \phi^{X}) }\\
\end{array}\right \}
\label{dictionaryR}
\end{equation}
The bosons $\phi^{ch},~\phi^{sp},~\phi^{fl}$ give rise to
the SU(4) group. The fifteen generators of SU(4) can be written
as
\begin{equation}
i\partial \phi^{ch} ~~~~~~~
i\partial \phi^{fl}~~~~~~~~
i\partial \phi^{sp}~~~~~~~~~
e^{\pm i \phi^{ch} \pm i \phi^{fl} }~~~~~~~~~
e^{\pm i \phi^{ch} \pm i \phi^{sp} }~~~~~~~~~
e^{\pm i \phi^{sp} \pm i \phi^{fl} } \nonumber
\end{equation}
All these generators should be conserved at the boundary so
\begin{equation}
\phi_L^{ch}(z) = \phi_R^{ch}({\bar z})|_{Im z =0}~~~~~~~~
\phi_L^{sp}(z) = \phi_R^{sp}({\bar z})|_{Im z =0}~~~~~~~~~~
\phi_L^{fl}(z) = \phi_R^{fl}({\bar z})|_{Im z =0}
\end{equation}
In terms of the fermions \ref{newfermions} these conditions
 can be rewritten,
together with the $\phi^X$ condition \ref{boundaryx} as
\begin{equation}
\psi_{L,ch}= \psi_{R,ch} \qquad \psi_{L,fl}= \psi_{R,fl}
\qquad \psi_{L,sp}=  \psi_{R,sp} \qquad
\psi_{L,x}=  \psi_{R}^{\dagger,x}
\label{monopoleboundary}
\end{equation}
As the sign of just one Majorana fermion is changed, we discover a
 hidden SO(7) symmetry in this theory. This symmetry is present
only in the case where the $\theta$ angle is zero. A non zero
$\theta$ angle changes the boundary condition for $\phi^X $ to
\begin{equation}
\phi^X_L(z) = -\phi^X_R({\bar z}) + \theta |_{Im z =0}
\end{equation}
which breaks SO(7) down to SO(6)$\sim$SU(4) again.

The boundary state and the partition function are described in
Appendix B, they agree with the ones found in \cite{affleck-saggi}.
We see that the only difference between the Kondo boundary condition
\ref{kondoboundary} and the Monopole \ref{monopoleboundary} is the
minus sign   for the fermion $\psi_{fl}$. This implies a
relationship  between the correlation functions of the two theories which
is described in Appendix A

\subsection{Two impurity Kondo problem}

We will first establish the connection between the description used in
 ref.\cite{ludwig-affleck-twoimp} in terms of
SU(2)$_1^{charge1}\times$SU(2)$_1^{charge2}\times$SU(2)$^{spin}_2\times$Ising
and  our four fermions \ref{newfermions}.
The two SU(2)$_1$ groups can be represented in terms
of two bosons $\phi^1,~\phi^2$ at the SU(2) compactification  radius.
They can be constructed  as the following linear combinations
$$ \phi^{1} = {\phi^{ch} + \phi^{fl} \over \sqrt{2} }~~~~~~~~~~~~~~~~~~~~~~~
\phi^{2} = {\phi^{ch} - \phi^{fl} \over \sqrt{2} }  $$
The Ising fermion can be identified with the Majorana fermion $\chi^X_1$
from \ref{majorana}.
The  SU(2)$^{spin}_2$ has the same representation as before
\ref{currentspmspin}.
The boundary condition in this model corresponds to flipping
the sign of the Ising fermion \cite{ludwig-affleck-twoimp}
 $$ \chi^X_{L1} = - \chi^X_{R1}$$
We see that this is the same boundary condition as in
\ref{monopoleboundary} except for an interchange $ \chi^X_1
\leftrightarrow \chi^X_2 $ so that
the two are
the {it same } conformal field theory. For example,
the partition functions are the same. They were
calculated independently in \cite{affleck-saggi} and
in \cite{ludwig-affleck-twoimp}.
We can also see from this boundary condition
the SO(7) symmetry of the theory, which is indeed
 the same group as in the
Monopole fixed point.

\section{Correlation functions for the two channel Kondo theory}

\subsection{Correlation function for the two channel Kondo effect
at the fixed point}

Using the bosonized picture \ref{newspinorsvector} together with the
boundary conditions for the bosons \ref{boundaryx}, \ref{boundaryspfl} and
\ref{boundarych} we can calculate all correlation functions of the theory
in a simple  fashion. As an example we will calculate the following
correlation function
\begin{eqnarray}
 & \langle \psi_{L11}(z_1) \psi^{\dagger 21}_L(z_2)
 \psi^{\dagger 12 }_R( \bar{z}_3 ) \psi_{R 22}(\bar{z}_4) \rangle
=
 \nonumber \\
& \langle
e^{-{i \over 2}( \phi_R^{ch}+ \phi_R^{sp}+ \phi_R^{fl}+ \phi_R^{X})(z_1) }
e^{{i \over 2}( \phi_R^{ch}- \phi_R^{sp}+ \phi_R^{fl}- \phi_R^{X})(z_2) }
e^{{i \over 2}
( \phi_R^{ch}+ \phi_R^{sp}+\pi- \phi_R^{fl}+ \phi_R^{X})( \bar{z}_3 ) }
\times ~~~~~~~~~~~~~~~~
\nonumber \\
&~~~~~~~~~~~~~~~~~~~~~e^{-{i \over 2}
( \phi_R^{ch}- \phi_R^{sp}-\pi- \phi_R^{fl}- \phi_R^{X})(\bar{z}_4) }
\rangle
= { -1 \over
[ (z_1-\bar{z}_3)(z_2-\bar{z}_3)
(z_1-\bar{z}_4)(z_2-\bar{z}_4) ]^{1\over 2}  }
\end{eqnarray}

In a similar fashion we can calculate all others. Strictly speaking we
should include cocycle operators that ensure proper fermionic
commutation relations for the exponentials. However, it can be shown that
they are not necessary
if we put the fermions first in some standard order and then replace them
by the exponentials.

We will calculate the four point functions of fermions and spin operators
for the two channel Kondo problem. We will show that all four point correlation
functions can be related to the one having four fermions. This is so because,
due to the Kondo boundary condition, two of these fermions turn into spin
operators. So no new calculations are necessary.

We are interested in the case were we have two right moving operators and
two left moving operators.
We will use the boundary condition \ref{kondoboundfer} to express the right
movers in terms of left movers analytically continued past the impurity.
The correlation function of  four fermions was analyzed in
\cite{ludwig-affleck-green}.
They calculated
\be \langle \p1 \psi^{\dagger \bb \jb}_L(z_2) \psi_{R\b j }
(\bar{z}_3) \psi_R^{\dagger \ab \ib}
(\bar{z}_4) \rangle_K = e^{-i\pi (j_\beta + j_{\ab} ) } \langle \p1
\psi^{\dagger \bb \jb}_L(z_2)
 S_{L\b j} (\bar{z}_3) S_L^{\dagger \ab \ib} (\bar{z}_4) \rangle_F
\label{oneandone} \ee
in formula (4.11) of that paper.

Now let us consider correlation functions with  spin operators. We will show
that
those cases can be put in a form similar to the one above.
The case of two left moving fermions and two right moving spin operators
reduces
simply to a trivial free fermion four point correlation function.

The nontrivial cases
 are when we have a left moving fermion and spin operator and similarly
for the right movers. Let us start with
$$ \langle \p1 S_{L \b j}(z_2) \psi^{\dagger \ab \ib}_R(\bar{z}_3 )
S_R^{\dagger \bb\jb}
(\bar{z}_4) \rangle_K = e^{-i \pi ( j_{\ab} + j_{\bb} ) } \langle \p1  S_{L \b
j}(z_2)
 S^{\dagger \ab \ib}_L(\bar{z}_3 ) \psi_L^{\dagger \bb\jb}
(\bar{z}_4) \rangle_F $$
We see that we can read the result for this correlation function from
\ref{oneandone}
making in that formula the replacements $z_2 \ra \bar{z}_4$, $\bar{z}_3
 \ra z_2$, $\bar{z}_4 \ra
\bar{z}_3 $. The formula has ambiguities of factors of $-1$ but those are
present because
there is a square root branch cut in the correlation function of the fermion in
the presence of a spin operator.
Another correlation function is
$$ \langle \p1 S_L^{\dagger \ab \ib}(z_2) \psi^{\dagger \bb \jb}_R (\bar{z}_3)
S_{R\b j}(\bar{z}_4) \rangle_K = e^{-i \pi (j_{\bb} + j_\beta )}
 \langle  \p1 S_L^{\dagger \ab \ib}(z_2)
S^{\dagger \bb \jb}_L (\bar{z}_3)\psi_{L\b j}(\bar{z}_4)  \rangle_F
$$
The result can be obtained from \ref{oneandone} by making the replacements
$\bar{z}_2 \ra \bar{z}_3$,
$z_3 \ra \bar{z}_4 $, $ \bar{z}_4 \ra z_2$. So that we get
$$
\langle \p1 S_L^{\dagger \ab \ib}(z_2) \psi^{\dagger \bb \jb}_R (\bar{z}_3)
S_{R\b j}(\bar{z}_4) \rangle_K=
{
e^{i\pi (j_{\ab} -j_\beta ) }
 ( \delta^{\ab}_\a \delta^{\bb}_\b \delta^{jb}_i \delta^{\ib}_j  -
\delta^{\bb}_\a
\delta^{\ab}_\b \delta^{\ib}_i \delta^{\jb}_j )  \over
[ (z_1 -\bar{z}_3)(z_1 - z_2) (\bar{z}_4 - \bar{z}_3) (\bar{z}_4 - z_2) ]^{1/2}
 }
$$
In this way we see that these correlation functions have the same form that
the ones for four fermions.
Let us remark that with this bosonized formalism it is also straightforward
to calculate higher point correlation functions

\subsection{ Correlation functions away from the fixed point}

This formalism enables us to calculate low temperature corrections
 to the scaling
behavior. They are  obtained adding to the action the leading irrelevant
operator,
which is the
boundary operator with lowest possible dimension that respects
all the symmetries of the problem. For our case this operator is
\cite{ludwig-fusion}
$ \phi_0 = \vec{J}^{spin}_{-1} \vec{\phi}^{spin} $
where $\vec{\phi}^{spin}$ is a charge zero, flavor-spin zero and
spin one primary field (the quantum numbers are $(Q, j^{spin},
j^{flavor} ) = (0,1,0)) $. The correction
term is
$$ S_{irr} = \lambda_0 \int dt i \chi_1 \chi_2 \chi_3   $$
where we have expressed $ \phi_0$ in terms of
 the three Majorana fermions associated with the
SU(2)$^{spin}_2$ group. The coupling $\lambda_0 $ has dimension 1/2 and
is proportional to $ T^{1\over 2} $.
In terms of the bosons \ref{newfermions} the correction to the action
  becomes
$$
 S_{irr} = -\lambda_0 \sqrt{2} \int dt \partial \phi^{sp} \sin \phi^X
$$

We saw that the left-right fermion green function vanishes
at the fixed point. So it is specially interesting to
calculate the leading irrelevant correction
since it is the dominant contribution.

The first order correction is  given by
$$ \langle \psi_{R++}^{\dagger }(\bar{z}') (-S_{int} ) \psi_{L++}(z)
\rangle = (-i) \lambda_0 \sqrt{2} \int dt \langle e^{ \frac{i}{2}
(-\phi^{ch} -\phi^{sp} - \phi^{fl} + \phi^X ) } \partial \phi^{sp} \sin
\phi^X e^{ \frac{i}{2} (\phi^{ch} + \phi^{sp} +\phi^{fl} + \phi^X )}
\rangle= $$
$$ = -i \lambda_0 2 \sqrt{2} (\bar{z}' -z)^{-3/2} $$

\section{Symmetry groups and boundary conditions}

The bulk Hamiltonian for eight Majorana fermions has an
O(8) symmetry. The fermions transform in the
vector representation and the two SO(8) spinors  form a
single 16 dimensional O(8) spinor. We saw above that by using the triality
transformation we generate an SO(8) group under which the
fermions are in a spinor representation. If we enlarge this
group to O(8)
 the original fermions together with the
$(s)$ spinor \ref{newspinorsvector} form a single 16 dimensional
representation which will be denoted by $\eta = (\psi_{\alpha j}, s_\mu)$.
 The extra group elements in O(8)
 that do not belong to SO(8)
 interchange the original
fermions with the $(s)$ spinor. Note that these elements
act locally on the fermions \ref{newfermions}.
The bulk free fermion theory is invariant under the transformations
\be
\eta_L = G_L \eta_L ~~~~~~~~~~~~~~~~~~\eta_R = G_R \eta_R
\label{transbulk}
\ee
where $G_L$ and $G_R$ are two independent group elements of O(8).
If we have a boundary condition of the form
\be
\eta_L = R \eta_R |_{Im z =0}
\label{bceta}
\ee
with R$\in$O(8)
then the transformations in \ref{transbulk} have to satisfy
\be
G_R = R^{-1} G_L R
\label{left-right}
\ee
Notice that this equation implies that if $G_L \in$ SO(8) then also
$G_R \in$ SO(8) for any $R$ in O(8). This implies that if we have a
theory with only the fermions (no $(s)$ spinor) then these
transformations leave them as fermions.
 In the problems we treated in this paper the boundary condition
can be put in this form with
R $\not\in$ SO(8) which means that the boundary condition \ref{bceta}
interchanges the fermions with the $(s)$ spinor.

For the free fermion theory R=1 so that $ G_L =G_R $ but for a general
group element
R we see from \ref{left-right}
that $ G_L =G_R $ implies that  $G_L$ commutes with R. We will call the
set of
such transformations the ``proper'' symmetry group of the theory.
If we allow different transformation laws for left and right movers
then the symmetry group of all these theories is always the same.
Instead, ``proper'' symmetries correspond to charges that
are conserved at the boundary.
We can determine $R$ for the different theories by its action
on the $(c)$ spinor
which is in
the vector representation of O(8). In the Monopole
theory
we see from
\ref{monopoleboundary} that $R_M$ can be represented by a diagonal matrix
that differs from the identity matrix only by the sign of one entry. This
implies that the ``proper'' symmetry group in
this case  is $Z_2\times$O(7) which reduces to the SO(7) we had before if
we consider only the elements continuosly connected with the identity.
The same situation is true for the two impurity Kondo effect since
it is the same conformal field theory as the one describing the
Monopole.

In the Kondo problem we see from \ref{kondoboundary} that $R_K$ differs
from the identity by the sign of three entries so that the proper group
 is O(3)$\times$O(5). This again reduces to SO(3)$\times$SO(5) if we
consider the part connected with the identity.

The correlation functions of any theory with a boundary condition
of the form \ref{bceta} can be obtained from the bulk free
theory by simply replacing $ \eta^a_L(z) \ra R^a_b  \eta^b_R( z)$ so that
we have
\begin{eqnarray}
\langle \eta^{a_1}_L(z_1)...\eta^{a_n}_L(z_n)
\eta^{c_1}_R(\bar{w}_1)...\eta^{c_m}_R(\bar{w}_m)
\rangle_{boundary} = \nonumber \\
 R^{a_1}_{b_1}...R^{a_n}_{b_n}
\langle  \eta^{b_1}_R(z_1)...\eta^{b_n}_R(z_n)
\eta^{c_1}_R(\bar{w}_1)...\eta^{c_m}_R(\bar{w}_m)
\rangle_{bulk}
\label{correlation}
\end{eqnarray}
Notice that if we want to calculate a correlation function
for left and right moving
fermions in the Kondo or Monopole model we have to know the
correlation function in the bulk for fermions {\it and spinors}, which
is a simple problem in free field theory once we use the bosonized
forms \ref{newspinorsvector}.

This implies also  a linear relationship between the correlation
functions of the Monopole and Kondo theories. These models are
specified by the group elements $R_M$ and $R_K$ appearing in the
boundary condition. Defining
$ S = R_K R^{-1}_M $ we have the following relation
\begin{eqnarray}
\langle \psi_L(z_1)...\psi_L(z_n) \psi_R(\bar{w}_1)...\psi_R(\bar{w}_m)
\rangle_{Kondo} =  \nonumber \\
\langle S \psi_R(z_1)... S \psi_R(z_n)
\psi_R(\bar{w}_1)...\psi_R(\bar{w}_m)
\rangle_{Monopole}
\end{eqnarray}

In appendix A we work out  this correspondence more explicitly.

\section{Conclusions}

We gave in this paper a unified description of several impurity
problems. The description was based in using the symmetry group of
the free theory SO(8). We used the triality transformation and we
enlarged the set of states to include the $(s)$ spinor. In this language
the boundary condition is linear and corresponds to an element of O(8).
We have solved the unitarity paradox by realizing that the fermions
are scattered to the $(s)$ spinor. We found the relation between the
correlation functions of the two channel Kondo effect and the four flavor
Monopole theory. We also established that the two impurity Kondo theory
is the same as the four flavor  Monopole theory.
We have reduced the problem of calculating correlation functions to
a free field theory exercise.

\centerline{\bf Acknowledgements}

We  thank C. Callan
 for useful discussions and comments.
We benefited also from  discussions with A. Yegulalp, I. Klebanov
and A. Aligia.

\appendix

\section{Mapping between the Kondo and Monopole correlators }

We remarked above that the Monopole and the Kondo correlation  functions
are linearly related. To make this more explicit we define
\begin{equation}
\psi^K_\mu =
\left \{
\begin{array}{l}
\psi^{\dagger,1,1} \\
\psi^{\dagger,2,1} \\
\psi^{\dagger,1,2} \\
\psi^{\dagger,2,2} \\
\psi_{1,1} \\
\psi_{2,1} \\
\psi_{1,2} \\
\psi_{2,2}
\end{array}
\right \}~~~~~~~~~~~~~~~~~~
\psi^M_\mu =
\left \{
\begin{array}{l}
\Psi^{\dagger,1} \\
 \Psi_{2} \\
 \Psi_{3} \\
\Psi^{\dagger,4} \\
\Psi_{1} \\
\Psi^{\dagger,2} \\
 \Psi^{\dagger,3} \\
 \Psi_{4} \\
\end{array}
\right \}
\end{equation}
and  we set them equal for right movers $\psi^K_R = \psi^M_R $.
We know that the difference between the Kondo and the Monopole
boundary conditions is just the phase associated with the third component
of the spin \ref{boundaryspfl}.
So for the left movers  we have
$ \psi^M_L = S \psi^K_L $ where $S=e^{-i\pi J_0^{sp}} $ is simply a phase,
different for different components of $\psi_\mu$. We have set the
 $\theta $ angle to zero for simplicity.
Using this we can obtain the Monopole correlation functions in terms of the
Kondo correlators. For example
\begin{equation}
\langle \Psi_1(z_1) \Psi_2(z_2) \Psi_3(\bar{z}_3)
\Psi_4(\bar{z}_4) \rangle = -
\langle \psi_{L11}(z_1) \psi^{\dagger 21}_L(z_2)
 \psi^{\dagger 12}_R( \bar{z}_3 )
\psi_{R 22}(\bar{z}_4) \rangle
\end{equation}
where the factor $-1$ came from the phase in the boundary condition.
All correlation functions of the Monopole theory can be mapped
on a correlation function of the Kondo theory.
We have checked explicitly that this is indeed true for all four point
functions which were calculated independently and
 using different
methods in \cite{ludwig-affleck-green} for the Kondo model and
in \cite{affleck-saggi} for the Monopole theory.

\section{Boundary states and partition functions}

We start by constructing the free fermion boundary state. It is
convenient to work in terms of the four bosons $\phi^{ch},\phi^{sp},
\phi^{fl}, \phi^X $. We denote their four  U(1) charges collectively
by $ \vec{J} =( J_0^{ch},J^{sp}_0,J^{fl}_0,J_0^X) $ and
 their eigenvalues by the four component vector $\vec{k}$.
The components of $\vec{k}$ are either all integer or all half integer.
In the Neveu-Schwartz sector $\sum_{i=1}^4 k_i =$ even, and in the Ramond
sector
this sum is odd.

The free fermion boundary state is
\be
\ket F = \exp ( \sum_{i=1}^4 { \alpha_{-n}^{L\,i} \alpha_{-n}^{R\,i} \over n} )
 \sum_{\vec{k} \in \Lambda}
 \ket{ {\vec{k}} }_L \ket{ - {\vec{k}} }_R
\label{freefermionbs}
\ee
where $ \alpha_{-n}^{L\,i} $, $\alpha_{-n}^{R\,i} $ are the left and
 right moving  oscillatory modes of the four
bosons and $\vec k $ are their momenta defined on
a lattice $\Lambda$ which depends on the sector we are considering.

 From the boundary conditions on the bosons  \ref{boundaryx}
\ref{boundaryspfl} \ref{boundarych} we conclude that the Kondo boundary
state is
\begin{eqnarray}
\ket K  = e^{i \pi  J^{sp}_L } {\cal R}_L
\ket F =~~~~~~~~~~~~~~~~~~~~~~~~~~~~~~~~~~~~~~~ \nonumber
\\
\exp ( \sum_{i=1}^3 { \alpha_{-n}^{L\,i} \alpha_{-n}^{R\,i} \over n} -
{ \alpha_{-n}^{L4} \alpha_{-n}^{R4} \over n}  )
 \sum_{\vec{k} \in \Lambda}
 e^{i \pi k_2} \ket{ (k_1,k_2,k_3,- k_4) }_L
\ket{ - (k_1,k_2,k_3, k_4) }_R
\label{kondobs}
\end{eqnarray}
where ${\cal R}_L$ reverses the left part of the boson $\phi^X$.

Using these boundary states it is easy to calculate the different
partition functions of the theory.
Actually without much more effort we can insert in the partition
function the four conserved charges. This will enable us to probe
in greater detail the quantum numbers of the states that are
propagating in each case.  We introduce for this purpose
 an angle for each charge $\vec{\varphi}_L = (\varphi^{ch}, \varphi^{sp},
\varphi^{fl}, \varphi^X)$ which will characterize the twist in the left sector
and a similar set of angles $\vec{\varphi}_R$ for the right sector.
The free fermion partition function is
\begin{equation}
Z_{FF} = \bra F q^{L_0 + \tilde{L}_0} e^{i\vec{\varphi}_L . \vec{J_L}}
e^{i\vec{\varphi}_R . \vec{J_R}} \ket F =
\bra F q^{L_0 + \tilde{L}_0}
e^{i(\vec{\varphi}_L-\vec{\varphi}_R ).\vec{J_L} }
\ket F
\end{equation}
with $q =e^{2 \pi l / \beta}$.
We used the conservation of the currents in the fermionic boundary state
to show that  the partition function will depend on $\vec{\alpha} \equiv
 \vec{\varphi}_L - \vec{\varphi}_R  $.

We will calculate the partition function for the NS sector
so we add over all $\vec k$  inserting the projection factor
 $ ( 1 +(-1)^{k_1 +k_2 +k_3 +k_4} ) \over 2 $. We obtain
\begin{eqnarray}
Z^{NS }_{FF} &=&
 {1\over 2} { \prod_{i=1}^4 ( \sum (q^2)^{n^2/2} e^{i n \alpha_i} ) +
 \prod_{i=1}^4 ( \sum (q^2)^{(n+1/2)^2/2} e^{i (n+1/2) \alpha_i} )
\over  (q^2)^{4/24} f(q^2)^4  }
\nonumber \\
&+& {1\over 2}
{ \prod_{i=1}^4 ( \sum (q^2)^{n^2/2} e^{i n (\alpha_i +\pi)} ) +
 \prod_{i=1}^4 ( \sum (q^2)^{(n+1/2)^2/2} e^{i (n+1/2) (\alpha_i +\pi)} )
\over  (q^2)^{4/24} f(q^2)^4  }
\label{ffpartition}
\end{eqnarray}

Now let us compute the partition function with a Kondo boundary
$$ Z^{NS}_{KF} = \bra K q^{L_0 + \tilde{L}_0}
e^{i\vec{\alpha}.\vec{J_L} } \ket F =\bra F {\cal R}
e^{- i \pi  J^{sp}}
 q^{L_0 + \tilde{L}_0} e^{i \vec{\alpha}.\vec{J_L} }  \ket F $$
The phase $e^{- i \pi  J^{sp}}$ can be absorbed redefining
$\alpha_2 \rightarrow \alpha'_2 =\alpha_2 - \pi$ and
$\alpha'_i =\alpha_i$ for $i\not=2$. The twist  in the $\phi^X $ boson
 leaves only the states with $k_4 =0$ and will produce a factor
of $1/\prod( 1+ q^{2n})$ from the oscillator modes. The result is
\begin{equation}
Z^{NS }_{KF} = {1\over 2} { \prod_{i=1}^3 ( \sum (q^2)^{n^2/2}
 e^{i n \alpha'_i} ) +
 \prod_{i=1}^3 ( \sum (q^2)^{n^2/2} e^{i n (\alpha'_i +\pi)} )
\over  (q^2)^{4/24} f(q^2)^3 \prod( 1+ q^{2n} )  }
\end{equation}
Setting $\vec{\alpha} =0$ and putting the result in terms of
$w = e^{-\pi \beta /l} $ we obtain
$$
 Z^{NS }_{KF} = {1\over 2} {
 \sum w^{n^2 \over 2} \sum w^{  {1\over 2} (n+{1 \over 2 } )^2 }
\sum w^{n^2 \over 8} \over w^{5/48} f(w)^3 \prod (1-w^{n-1/2}) }
$$
Which agrees with the result obtained in \cite{ludwig-fusion}
 using the fusion hypothesis.
In the case where we have the Kondo boundary condition on
both ends  we could compute $\bra K  q^{L_0 + \tilde{L}_0}
e^{i ( \vec{\varphi}_L .\vec{J_L} - \vec{\varphi}_R .\vec{J_R} ) }
\ket K $. This gives the same result as
the free fermion partion function
but with $\alpha_4 \rightarrow \hat{\alpha}_4 =
 \varphi_{4L} +\varphi_{4R} $ and $\hat{\alpha}_i = \alpha_i$ for i=1,2,3.
This reflects the change in the U(1)$^X$ boson boundary condition. Note
that if we set all angles to zero we obtain exactly the same result, the
angles enable us to distinguish between the two situations. This partition
function however does not agree with the one calculated using fusion because
we have here states propagating that are in the (R,NS) sector. If
we restrict just to states in the (NS,NS) sector we obtain
\begin{equation}
Z^{NS }_{KK} = {1\over 2} { \prod_{i=1}^4 ( \sum (q^2)^{n^2/2}
 e^{i n \hat{\alpha}_i} ) +
 \prod_{i=1}^4 ( \sum (q^2)^{n^2/2} e^{i n (\hat{\alpha}_i +\pi)} )
\over (q^2)^{4/24} f(q^2)^4  }
\end{equation}
which is indeed the result that the fusion hypothesis produces.

We can also repeat the same calculations including, in the
free fermion boundary state, also states in the R sector. In the
open string picture this would correspond to the NS sector with
only states of even fermion number. The calculations are
very similar to the ones outlined here, in this case we do not
project on states for which the sum of $k_i$ is even. The results
agree with the ones that one could calculate using fusion. And,
again, when both boundaries are Kondo we have to project on the
sectors (NS,NS) and (R,R).

Now we consider the Monopole partition functions. The Monopole
boundary state is just
\be
\ket{ M \theta}  =  e^{i \theta J_L^{X} } {\cal R } \ket F
\ee
where we have included also the effect of a non vanishing $\theta$ angle.
We can write the partition function immediately from the
expressions we had for the Kondo case. We want however to
make connection with the description using SU(4) symmetry
so let us analyze the SU(4) characters first.
 The SU(4) group has three Cartan generators that
we can associate to the bosons $(\partial
\phi_1,\partial \phi_2,\partial \phi_3 )$
while
the other twelve generators are
given by $e^{i (\pm \phi_i \pm \phi_j)}$ (with $ i \not= j $).
The Kac-Moody algebra
 SU(4)$_1$ has four representations: $ 1, 4, \bar{4}, 6$,
(we labeled them according to their dimensions) of weights:
$0,{3\over 8},{3\over 8},{1\over 2}$
 respectively.
The states  have charges characterized by $ e^{i(k_1 \phi_1 +k_2 \phi_2 +
k_3 \phi_3)} $
with $k_i$ all integer or all half integer. The representation to which this
 state belongs
depends on the sum of $k_i$
$$ k_1 +k_2 + k_3 = \left\{ \begin{array}{ll} \mbox{even}
&~~~~~(1) \\
\mbox{odd} &~~~~~(6) \\ 3/2 + \mbox{even} &~~~~~(4) \\
3/2 +\mbox{odd} &~~~~~(\bar{4}) \end{array}
\right.
$$
The SU(4)  non specialized
 characters are obtained inserting  an exponential of the form
$e^{\sum \alpha_i j_i}$
where $j_i$ are the Cartan generators of SU(4). With the conventions
$\vec{n} = (n_1,n_2,n_3)$, $\ve2=({1\over 2},{1\over 2},{1\over 2}) $ and
$\vec{\alpha}=(\alpha_1,\alpha_2,\alpha_3)$
they read
\begin{eqnarray}
\chi^{SU(4)}_1(q,\vec{\alpha}) &=&
 {1\over 2} { \sum_{\vec{n}} (q)^{{\vec{n}}^2/2}
e^{i \vec{n}.\vec{\alpha}}(1 + (-1)^{n_1 + n_2+ n_3} ) \over
q ^{3/24} f(q)^3 }
\\
\chi^{SU(4)}_6(q,\vec{\alpha}) &= &
{1\over 2} { \sum_{\vec{n}} (q)^{{\vec{n}}^2/2 }
e^{i\vec{n}.\vec{\alpha}}(1 - (-1)^{n_1 + n_2+ n_3} ) \over
q ^{3/24}f(q)^3 }
\\
\chi^{SU(4)}_4(q,\vec{\alpha}) &=&
 {1\over 2} { \sum_{\vec{n}} (q)^{{(\vec{n}+\ve2)}^2/2\ }
e^{i(\vec{n}+\ve2).\vec{\alpha}}
(1 + (-1)^{n_1 + n_2+ n_3} ) \over
q ^{3/24}f(q)^3 }
\\
\chi^{SU(4)}_{\bar{4}}(q,\vec{\alpha}) &=&
 {1\over 2} { \sum_{\vec{n}} (q)^{{(\vec{n}+\ve2)}^2/2 }
e^{i(\vec{n}+\ve2).\vec{\alpha}}
(1 - (-1)^{n_1 + n_2+ n_3} ) \over
q ^{3/24}f(q)^3 }
\end{eqnarray}

Now we analyze the U(1)  characters. The non
specialized characters  for the different values of $Q$
( $Q =\, 0,\pm 1/2,\,1 $) are
\begin{equation}
\chi^{{\cal A}_4}_Q(q,\alpha) = {\sum q^{ (Q + 2 n)^2/2 }
e^{i (Q + 2 n) \alpha }
\over q^{1/24} f(q) }
\end{equation}
We find that the partition function \ref{ffpartition} can be rewritten as
\begin{eqnarray}
Z_{FF}^{NS}(q^2,\alpha_1,\alpha_2,\alpha_3,\alpha_4)
& = &\chi^{U(1)}_0(q^2,\alpha_4) \chi^{SU(4)}_{1}(q^2,\vec{\alpha})
+ \chi^{U(1)}_{1/2} (q^2,\alpha_4) \chi^{SU(4)}_{4}(q^2,\vec{\alpha})
\nonumber \\
&+& \chi^{U(1)}_{-1/2} (q^2,\alpha_4) \chi^{SU(4)}_{\bar{4}}(q^2,\vec{\alpha})
+ \chi^{U(1)}_1(q^2,\alpha_4) \chi^{SU(4)}_{6}(q^2,\vec{\alpha})
\end{eqnarray}
The partition function with the Monopole boundary conditions on one
side and free fermions on the other is
$$
Z_{M\theta ,F}^{NS} = { 1\over (q^2)^{1/24} \prod (1+q^{2n}) }
 \chi^{SU(4)}_{1}(q^2,\vec{\alpha})
$$
Notice that the relationship with the Kondo partition
function is very simple
$$
Z_{K,F}^{NS}(q^2,\alpha_1,\alpha_{2},\alpha_{3},\alpha_4) =
Z_{M,F}^{NS}(q^2,\alpha_1,\alpha_{2}-\pi,\alpha_{3},\alpha_4)
$$
The partition function with Monopole boundary conditions on both sides
and different theta angles is
$$
Z_{M\theta,M\theta'}^{NS}
=\chi^{U(1)}_0 (q^2,\tilde{\alpha}_4) \chi^{SU(4)}_{1}(q^2,\vec{\alpha})
+ \chi^{U(1)}_1 (q^2,\tilde{\alpha}_4) \chi^{SU(4)}_{6}(q^2,\vec{\alpha})
$$
where $\tilde{\alpha}_4 = \varphi_{4L} + \varphi_{4R} -\theta +
\theta' $ and
again we have projected on to (NS,NS) states.
When we set all $\varphi =0$ these expressions
agree with the ones in \cite{affleck-saggi}.


\begin{references}

\bibitem{cklm} C. Callan, I Klebanov, A. Ludwig and J. Maldacena,
Nucl. Phys. B422 (1994) 417

\bibitem{ludwig-swaves} I. Affleck and A. Ludwig, Nucl. Phys. B360, 641 (1991)

\bibitem{emery-kivelson} V. Emery and S. Kivelson Phys. Rev. B46 (1992) 10812

\bibitem{affleck-saggi} I. Affleck and J. Sagi, Nucl. Phys. .B417, 374 (1994)

\bibitem{ludwig-affleck-green} A. Ludwig and I. Affleck, Nucl.
Phys. B428 (1994) 545.

\bibitem{ludwig-fusion} I. Affleck and A.W.W. Ludwig, Nucl.Phys.
B 360 (1991) 651; A.W.W.Ludwig, Int. J. Mod. Phys. B8 (1994) 347.

\bibitem{cardy} J.L. Cardy, Nucl. Phys. B 324 (1989) 581.


\bibitem{ludwig-amplitude} I. Affleck and A. Ludwig Phys. Rev. B 48 (1993)
7297;
A.W.W. Ludwig and I.Affleck, Phys. Rev. Lett. 67 (1991) 3160.

\bibitem{callan-rubakov} V.A. Rubakov Nucl. Phys. B203 (1982) 311. C. Callan
Phys. Rev. D25 (1982) 2058; Nucl. Phys. B212 (1983) 391.

\bibitem{polchinsky} J. Polchinski
Nucl. Phys. B242 (1984) 345

\bibitem{ludwig-affleck-twoimp} I. Affleck, A. Ludwig and B. Jones
UBCTP-94-003/
PUPT-94

\bibitem{georgi-slansky}  see e.g: H. Georgi, ``Lie Algebras in
Particle Physics'' [Bejamin/Cummings, Menlo Park, 19982].

\end{references}
 \end{document}